\documentclass[12pt]{article}

\usepackage{epsfig}
\usepackage{graphicx}

\def\be{\begin{equation}}
\def\ee{\end{equation}}
\def\beq{\begin{eqnarray}}
\def\eeq{\end{eqnarray}}

\def\({\left (}
\def\){\right )}
\def\[{\left [}
\def\[{\right ]}

\begin{document}

\begin{titlepage}
\bigskip
\rightline{}
\rightline{hep-th/0409314}
\bigskip\bigskip\bigskip\bigskip
\centerline
{\Large \bf {Stability and Thermodynamics of }}
\bigskip
\centerline{\Large \bf {AdS Black Holes with Scalar Hair }}
\bigskip\bigskip
\bigskip\bigskip

\centerline{\large Thomas 
Hertog\footnote{Hertog@vulcan.physics.ucsb.edu} and 
Kengo Maeda\footnote{kmaeda@kobe-kosen.ac.jp}}
\bigskip\bigskip
\centerline{\em ${}^1$ Department of Physics, UCSB, Santa Barbara, CA 93106}
\bigskip
\centerline{\em ${}^2$ Department of General Education, Kobe City College 
of Technology, }
\centerline{\em 8-3 Gakuen-higashi-machi, Nishi-ku, Kobe 651-2194, Japan}
\bigskip\bigskip
\begin{abstract}

Recently a class of static spherical black hole solutions with scalar hair 
was found in four and five dimensional gauged supergravity with modified, 
but AdS invariant boundary conditions. These black holes are fully specified 
by a single conserved charge, namely their mass, which acquires a contribution
from the scalar field. Here we report on a more detailed study of some of the
properties of these solutions.  
A thermodynamic analysis shows that in the canonical ensemble the standard
Schwarzschild-AdS black hole is stable against decay into a hairy black hole.
We also study the stability of the hairy black holes and find there always
exists an unstable radial fluctuation, in both four and five dimensions. 
We argue, however, that Schwarzschild-AdS is probably not the 
endstate of evolution under this instability.

\end{abstract}

\end{titlepage}

\baselineskip=18pt

\setcounter{equation}{0}
\section{Introduction}

The original no hair theorem of Bekenstein \cite{Bekenstein74} proves there 
are no asymptotically flat black hole solutions with scalar hair for minimally
coupled scalar fields with convex potentials. This was generalized to 
minimally coupled scalars with arbitrary positive potentials in 
\cite{Heusler92}. But the no scalar hair theorem does not hold, in general, if
the scalar field asymptotically tends to a {\em local} extremum of the 
potential \cite{Sudarsky02}. In asymptotically flat space, this obviously 
requires potentials with negative regions. While potentials of this type 
arise generically as effective four dimensional potentials in Calabi-Yau 
compactifications \cite{Hertog03a}, the observation of \cite{Sudarsky02}
also leads to 
asymptotically anti-de Sitter (AdS) space as a natural context to study 
hairy black holes. 

Static, spherically symmetric asymptotically AdS black holes with scalar hair
were found (for minimal coupling) analytically in three dimensions 
\cite{Henneaux02}, and numerically in four 
\cite{Torii:2001pg,Winstanley03,Hertog04} and five \cite{Hertog04} dimensions.
In four dimensions, a class of hairy black holes that are asymptotically
locally AdS was found in \cite{Martinez04}. In most of these examples the 
scalar field asymptotically goes to a negative maximum of the potential, but
the scalar field mass is generally taken to satisfy the 
Breitenlohner-Freedman (BF) bound 
\cite{Breitenlohner82}. For certain boundary conditions, this ensures that the
AdS solution itself is nonlinearly stable provided the potential is of the form
$V=(D-2)W'^2 -(D-1)W^2$ in $D$ dimensions \cite{Townsend84}. 
However, the asymptotic conditions obeyed by all
known hairy black hole solutions do {\em not} belong to this class. 
Although they preserve AdS invariance \cite{Henneaux02,Hertog04,Henneaux04}, 
they render the AdS vacuum nonlinearly unstable \cite{Hertog04c}.

Nevertheless, the hairy black holes found in \cite{Hertog04}
are solutions of a consistent
truncation of ${\cal N}=8$ gauged supergravity in four and five dimensions.
These theories are thought to arise as the low energy limit of M/string
theory with boundary conditions $AdS_4 \times S^7$ and $AdS_5 \times S^5$.
For those boundary conditions we have the celebrated AdS/CFT correspondence
\cite{Maldacena98}, which provides a non-perturbative definition of
string theory on asymptotically AdS spacetimes in terms of a conformal field 
theory (CFT). In four dimensions, the 
hairy black holes obey unusual, but AdS invariant asymptotic conditions,
which deform the usual dual 2+1 CFT on a stack of M2 branes by
a triple trace operator \cite{Hertog04,Witten02}. The hairy black holes are 
described by approximately thermal states in this deformed dual field theory.
AdS/CFT relates the thermodynamics of AdS black holes to the 
thermodynamics of the dual CFT \cite{Witten98}, providing a novel way
to study phase transitions in field theories. Vice versa, one can 
hope to use the AdS/CFT correspondence to better understand the
microscopic description of the hairy black holes. In particular, it is an
interesting question how string theory distinguishes between a hairy black
hole and Schwarzschild-AdS of the same mass, which is a solution too 
for the same boundary conditions.

With these general motivations in mind, we now turn to a more detailed study 
of the hairy black holes found in \cite{Hertog04}.

\setcounter{equation}{0}
\section{Hairy black holes in $D=4$ Supergravity}

We consider four dimensional gravity minimally coupled to a scalar field 
with action
\be \label{act}
S=\int d^4x\sqrt{-g}\left[\frac{1}{2}R-\frac{1}{2}(\nabla\phi)^2 
+2 +\cosh (\sqrt{2} \phi)\right], 
\ee
where we have set $8\pi G=1$. 
This is a consistent truncation of ${\cal N}=8$ gauged supergravity in four
dimensions \cite{Duff99}, 
which is the low energy limit of string theory with 
$AdS_4 \times S^7$ boundary conditions. The potential has a global 
negative maximum
at $\phi=0$, where the scalar has mass (note that $l^2_{ads} =1$)
\be
m^2=-2.
\ee
This is slightly above the Breitenlohner-Freedman bound
\cite{Breitenlohner82}, 
\be
m^2_{BF} =\frac{-(D-1)^2}{4}, 
\ee
in $D=4$ dimensions, which ensures the perturbative 
stability of the $AdS$ solution. Here we are interested in
nonlinear perturbations of AdS. In particular we will consider the 
class of asymptotically anti-de Sitter 
solutions specified by the following set of asymptotic 
conditions 
\be 
\label{4-scalar}
\phi (r,t,x^{a})=\frac{\alpha(t,x^a)}{r}+
\frac{c\alpha^2(t,x^a)}{r^2}
\ee
\beq 
\label{4-grr}
g_{rr}=\frac{1}{r^2}-\frac{(1+\alpha^2/2)}{r^4}+ \frac{M_0(t,x^a)}{r^5} 
& \quad g_{tt}=-r^2 -1+O(1/r) \nonumber\\
g_{tr}=O(1/r^2) \qquad \qquad \qquad & \ \  \ \ \ g_{ab}= \bar g_{ab} +O(1/r) 
\nonumber\\
g_{ra} = O(1/r^2) \qquad \qquad \qquad & g_{ta}=O(1/r) \ \ 
\eeq
where $x^a=\theta,\phi$,  and $c$ is an arbitrary constant that labels 
a one-parameter class of different boundary conditions. For all values of $c$
this set of asymptotic conditions on the fields preserves the full AdS 
symmetry group \cite{Hertog04}, 
despite the fact that the asymptotic
behavior of certain metric components is generally relaxed, compared to the
standard set of AdS invariant boundary conditions \cite{Henneaux85}.
The conserved charges that generate the AdS symmetries 
are well-defined and finite for all $c$, but 
acquire a contribution from the scalar field.

Spherically symmetric solutions can be written as
\be \label{hairymet}
ds^2=-fe^{-2\delta}dt^2+f^{-1}dr^2+r^2d\Omega^2, \nonumber \\
\ee
where $f(t,r)$ and $\delta(t,r)$. The field equations reduce to
\be \label{field1}
(1-f) - rf'=r^2\left[\frac{1}{2}f^{-1}e^{2\delta}\dot{\phi}^2
+\frac{f}{2}{\phi'}^2+V(\phi)\right],  
\ee
\be \label{field2}
\dot{f}=-rf\dot{\phi}\phi', 
\ee
\be \label{field3}
2\delta'=-r[f^{-2}e^{2\delta}\dot{\phi}^2+\phi'^2], 
\ee
\be \label{field4}
-\dot{(e^\delta f^{-1}\dot{\phi})}
+\frac{1}{r^2}(r^2e^{-\delta}f\phi')'=e^{-\delta}\frac{dV}{d\phi},  
\ee
where $\dot \phi = \partial_{t}\phi$ and $\phi'=\partial_{r}\phi$.

In \cite{Hertog04} it was shown that the theory (\ref{act}) with 
boundary conditions (\ref{4-scalar})-(\ref{4-grr}) and $c \neq 0$
admits a one-parameter class of regular
static, spherically symmetric black hole solutions with scalar hair outside 
the horizon. We summarize the results of \cite{Hertog04} in 
Figure 1, where we plot the value $\phi_e$ of the field at the horizon as 
a function of horizon size $R_e$, for two different choices of 
AdS invariant boundary conditions, namely $c=-1$ (bottom) and 
$c=-1/4$ (top).  The hairy black hole solutions are specified by a single
conserved charge, their mass, which is given by
\be \label{mass4dhair}
E_{h} =4\pi \left( M_0+\frac{4}{3}c\alpha^3 \right),
\ee
where the second term is a finite scalar contribution. The mass was computed
in \cite{Hertog04}, where it was shown that $E_h \sim R_e^3$ for large $R_e$.
For a given horizon size $R_e$ one has 
$E_h > E_s = 4\pi(R_e+R_e^3)$. For large $R_e$ it was found
(see Figure 10 in \cite{Hertog04}) that $E_h \rightarrow E_s$, which means
$E_h = 4\pi R_e^3 (1+ {\cal O} (1/R_e))$ in this regime.
On the other hand, 
since Schwarzschild-AdS is a solution too for all $c$, one has 
two very different black hole solutions for a given mass. The hairy 
black holes thus provide an example of black hole non-uniqueness.

\begin{figure}[htb]
\begin{picture}(0,0)
\put(56,254){$\phi_e$}
\put(412,23){$R_{e}$}
\end{picture}
\mbox{\epsfxsize=14cm \epsfysize=9cm \epsffile{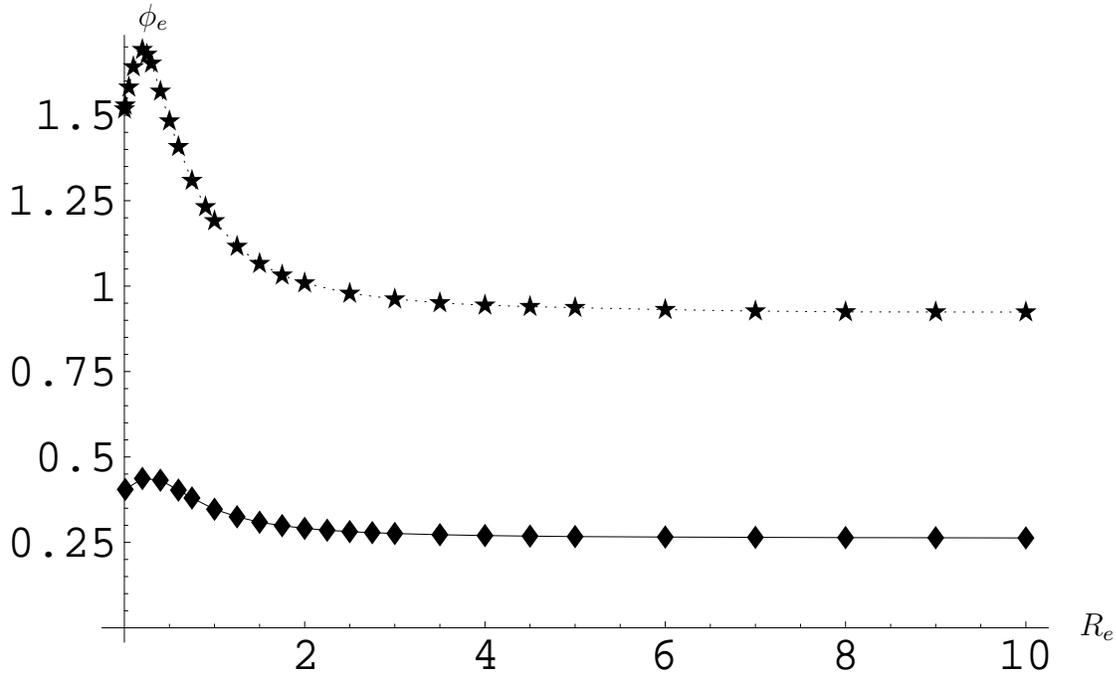}}
\caption{The scalar field $\phi_e$ at the horizon as a function of horizon size
$R_e$ in hairy black hole solutions of $D=4$ gauged supergravity.
The two curves correspond to solutions with two different AdS invariant 
boundary conditions, labelled by $c=-1$ (bottom) and $c=-1/4$ (top).}
\label{1}
\end{figure}

\setcounter{equation}{0}
\section{Stability}

We now consider linear radial fluctuations around the
hairy black hole background. Substituting 
\beq 
\phi(r,t)&=&\phi_0(r)+\phi_1(r)e^{i\omega t}\eta, \nonumber \\
f(r,t)&=&f_0(r)+f_1(r)e^{i\omega t}\eta, \nonumber \\
\delta(r,t)&=&\delta_0(r)+\delta_1(r)e^{i\omega t}\eta, 
\eeq
in the field equations (where
$(f_0(r),\delta_0(r),\phi_0(r))$ are the background quantities) 
and expanding in $\eta$ yields, to first order, the following perturbation 
equation for the scalar field,
\be 
\frac{e^{2\delta_0}}{f_0}\frac{d^2\phi_1}{dr^{*2}}
+\frac{2f_0}{r}\phi_1'
+\left[e^{2\delta_0}f_0^{-1}\omega^2-2f_0{\phi_0'}^2
-\frac{r^2}{2}f_0{\phi_0'}^4-(r\phi_0'f_0)'\right]\phi_1
=V_{,\phi \phi} (\phi_0) \phi_1,  
\ee
where $dr^*/dr=e^{\delta_0}/f_0$. 
Eliminating $\phi_0''$ using eq. (\ref{field4}) and defining 
$\chi=r\phi_1$, one obtains
\be \label{perteq}
-\frac{d^2\chi}{dr^{*2}}+P\chi=\omega^2\chi
\ee
with
\be
P=f_0 e^{-2\delta_0}\left(
-\frac{1}{2}({\phi'_0}^2+r^2{\phi'_0}^4)f_0+
\left(\frac{1}{r}-r{\phi'_0}^2\right)f'_0
+2r\phi'_0 V_{,\phi}(\phi_0)+V_{,\phi \phi}(\phi_0) \right). 
\ee
A hairy black hole is unstable if there exists a solution to 
(\ref{perteq}) with $\omega^2 <0$, since small
fluctuations of this kind grow exponentially in time. 

One immediately sees that at the horizon the potential $P$ vanishes. 
Near spacelike infinity on the other hand one has, for general $m^2$,
\be
P \rightarrow (m^2 -m^2_c) e^{-2\delta_0}r^2,
\ee
where $m^2_c=-D(D-2)/4$ is the conformal mass. Thus
$P$ is unbounded from below for scalars with mass $m^2 < m^2_{c}$.
The fact that the conformal mass separates two qualitatively different regimes
is true in any dimension. For $m^2 = m^2_c$, which is the case of
interest to us, the subleading terms are important at large $r$, giving
\be
P \rightarrow \frac{-\alpha^2 }{2} e^{-2\delta_0} \def \bar{p}.
\ee
Therefore, $P$ tends to a negative constant in our case, 
for all choices of boundary conditions. In Figure 2 we show the potential
for perturbations around a hairy black hole of size $R_e=2$, 
in two different theories, specified by $c=-1$ and $c=-1/4$.

\begin{figure}[htb]
\begin{picture}(0,0)
\put(27,125){$P$}
\put(204,80){$r$}
\put(250,141){$P$}
\put(455,20){$r$}
\end{picture}
\mbox{
\epsfxsize=7cm
\epsffile{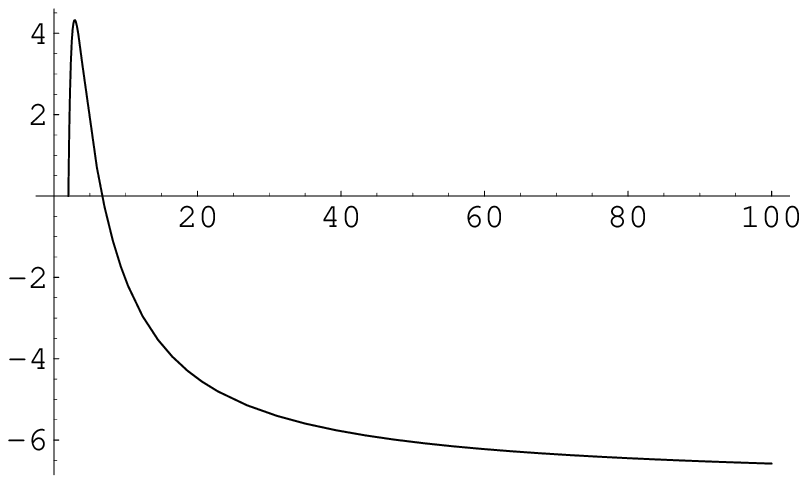}
\raisebox{2.3cm}{~~~~
\begin{minipage}{10cm}
\epsffile{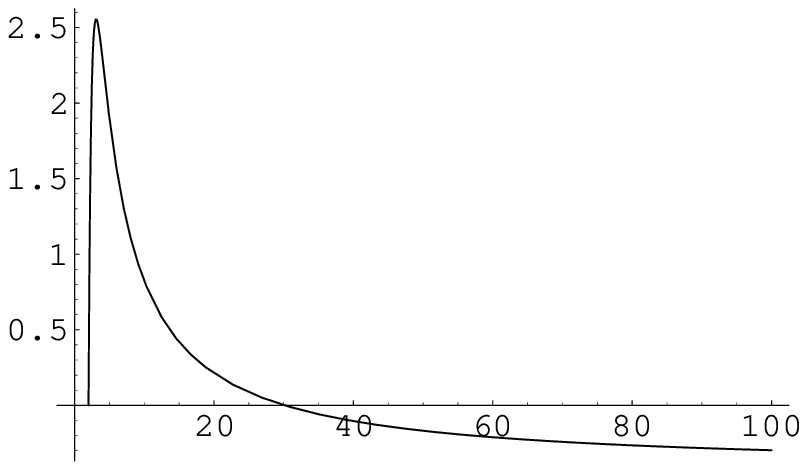}
\end{minipage}}}
\caption{Potential for linearized
radial fluctuations around a hairy black hole with radius $R_e =2$, 
for $c=-1/4$ (left) and $c=-1$ (right) boundary conditions.}
\label{2}
\end{figure}

The fact that $P$ is asymptotically negative means there exist exponentially 
growing fluctuations, potentially causing the hairy black holes to be unstable.
It only remains to verify there exists an unstable mode that obeys the 
modified boundary conditions with the same value of $c$ as the black hole 
background. Near the horizon we have $r^*\to -\infty$ and $P \to 0$, 
so the general near horizon solution of (\ref{perteq}) is given by 
\be \label{aschi}
\chi =  ae^{\sigma r^*} +be^{-\sigma r^*}, 
\ee
where $\omega=-i\sigma$. Regularity at the horizon requires $b=0$, 
since future unstable modes have $\sigma >0$. 
This translates into the following boundary condition
in terms of the original $r$ coordinate. 
From (\ref{field1}) and (\ref{field4}) it follows
\be 
\kappa\equiv {f_{,r}}|_{R_e}=
\frac{1-R_e^2V(\phi(R_e))}{R_e}.   
\ee
Since $f\simeq \kappa(r-R_e)$ near the horizon, 
this gives
\be 
r^*\simeq\frac{e^{\delta_0}}{\kappa}\ln (r-R_e)=
\frac{\ln (r-R_e)}{\kappa},   
\ee
where we have set $\delta_0(R_e)=0$. 
Hence (\ref{aschi}) becomes
\be 
\chi=a(r-R_e)^{\frac{\sigma}{\kappa}},  
\ee
giving, at $\epsilon = (r-R_e)$,
\be 
\frac{d\chi}{dr}(\epsilon)=\frac{\sigma \chi(\epsilon)}{\kappa \epsilon}
\ee

\begin{figure}[htb]
\begin{picture}(0,0)
\put(50,254){$\omega^2_{n}$}
\put(412,245){$R_{e}$}
\end{picture}
\mbox{\epsfxsize=14cm \epsfysize=9cm \epsffile{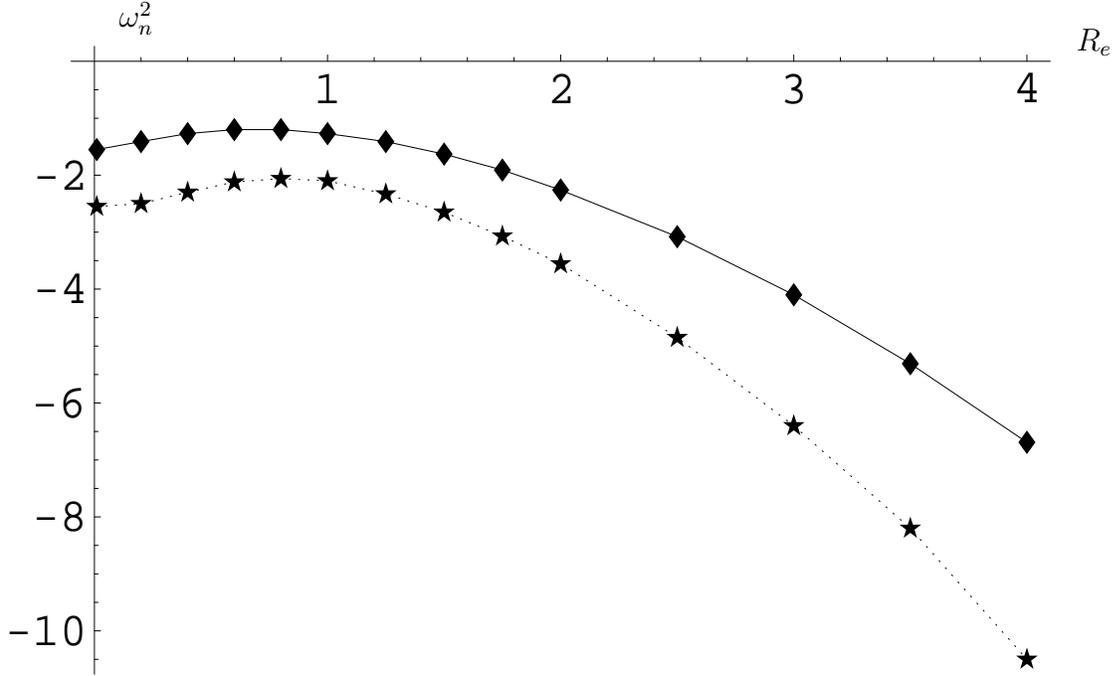}}
\caption{The value of the negative frequency, $\omega^2_{n}$,  
of the unstable radial perturbation as a function of horizon size $R_e$, 
in $D=4$ supergravity with
two different AdS invariant boundary conditions, namely $c=-1$ (bottom) and 
$c=-1/4$ (top).}
\label{3}
\end{figure}

Finally, to satisfy (\ref{4-scalar}) with the same value of $c$ as the background,
we must require that near spacelike infinity the fluctuations behave as
\be \label{ascondpert}
\phi_1(r)=\frac{\chi}{r} =
\left(\frac{\alpha_1}{r}+\frac{2c\alpha_0 \alpha_1}{r^2}\right). 
\ee
Here $\alpha_0$ is the coefficient of the $1/r$ mode of the background
solution, $\phi_0$, and $\alpha_1$ is an arbitrary constant.
Now, for large $r$, equation (\ref{perteq}) for the fluctuations
is approximately
\be 
-\frac{d^2\chi}{dr^{*2}}=(\omega^2-\bar{p})\chi=\tilde \omega^2\chi.
\ee
In the frequency regime of interest, $\bar{p}<\omega^2<0$, this yields
\be 
\chi=A\cos(\tilde \omega r^*+\gamma)\simeq 
A\cos(-\tilde \omega /r+k\tilde \omega+\gamma) \sim 
A\left(\cos(k\tilde \omega+\gamma)
+\frac{\tilde \omega}{r}\sin(k\tilde \omega+\gamma)\right), 
\ee
where we used $r^*\simeq -1/r+k$ with $k$ a constant. 
Therefore, the constraint (\ref{ascondpert}) translates into
the following condition on the frequency $\tilde \omega$,
\be \label{condfreq}
\tan(k\tilde \omega+\gamma)=\frac{2c\alpha_0}{\tilde \omega}. 
\ee

For given $\omega^2$, the boundary conditions at the horizon uniquely determine
the solution for the rescaled fluctuation $\chi(r)$. But we find that for all
horizon sizes $R_e$, there is precisely one negative frequency, 
$\omega^2_{n}$, 
for which the asymptotic condition (\ref{condfreq}) is satisfied too. 
Our results are shown in Figure 3, where we plot the value of this 
frequency, $\omega^2_{n}$, 
as a function of the black hole radius $R_e$, in two different theories.
For large $R_e$, we find $\omega^2_{n}  \propto R_e^2$, which is natural since
$ \bar{p} \propto \alpha^2 \propto R_e^2$ (the last step follows from the results 
in \cite{Hertog04}). Hence we 
conclude that the hairy black holes found in \cite{Hertog04} are unstable
to linearized radial fluctuations exponentially growing in time.

\setcounter{equation}{0}
\section{Thermodynamics}

The canonical ensemble is defined by a Euclidean path integral over matter 
fields and metrics which tend asymptotically respectively to zero and AdS 
space, identified periodically in $\tau =it$ with period $\beta$. The inverse 
of the period $\beta$ corresponds to the temperature $T$. The path integral 
is usually approximated by summing over saddle-points. Periodically identified
AdS space is one of these and we take it to be the zero of action and energy. 
This gives the dominant contribution at low temperature, but it is well known
\cite{Hawking83} that at sufficiently high temperature a large Euclidean 
Schwarzschild-AdS black hole solution has lower action. In this regime the pure
radiation will then tend to tunnel to the black hole configuration at a rate 
determined by the difference of the actions of both geometries.

Here we consider a canonical ensemble described by a path integral with 
boundary conditions given by (\ref{4-scalar}) and (\ref{4-grr}) for some $c$. 
The Euclidean continuation of the hairy black hole solution gives an extra 
saddle-point contribution. This raises the question whether there is a 
temperature regime where Schwarzschild-AdS is likely to `decay' into a hairy 
black hole, hereby spontaneously acquiring a nontrivial scalar field profile 
outside its horizon.

To study this issue it is sufficient to consider the minisuperspace consisting
of static, spherically symmetric Euclidean metrics, 
\be \label{mini}
ds^2=e^{-2\delta(r)}f(r)d\tau^2+f^{-1}(r)dr^2+r^2d\Omega^2,  
\ee
where $ 0 \leq t \leq \beta$, $r \geq R_e$ and the scalar field $\phi (r)$. 
The reduced Hamiltonian action reads,
\beq \label{eucact}
I &=& \int N{\cal H}+B\nonumber\\
&=& 4\pi\beta\int^\infty_{R_e}
 e^{-\delta} r^2\left[\frac{f'}{r}
-\frac{1}{r^2}(1-f)+\frac{1}{2}f\phi'^2+V(\phi)\right]dr+B,   
\eeq
where  
\be \label{varsurf}
\delta B=-4\pi\beta e^{-\delta}[r\delta f+r^2f\phi'\delta\phi]^\infty_{R_e}. 
\ee
The variation of the boundary term is determined by the condition
that the action is an extremum under the variation of the fields considered 
here \cite{Henneaux85,Regge74}. 
The last term in (\ref{varsurf}) comes from the scalar field variation. 
The geometries in the variation are smooth and complete if and only if the 
period $\beta$ satisfies  
\be \label{bet}
\beta e^{-\delta(R_e)}f'(R_e)=4\pi.
\ee
We first consider the contribution from the Euclidean 
continuation of the hairy black hole solution (\ref{hairymet}).
For this, (\ref{bet}) gives the temperature as a function of horizon 
size $R_e$,
\be
2\pi T = \frac{2\pi}{\beta} = \frac{1-V(\phi(R_e))R_e^2}{2R_e}. 
\ee
One sees that, like Schwarzschild-AdS, the hairy black hole only contributes 
to the thermodynamic ensemble at high temperatures, 
$2\pi T \ge \sqrt{-V(\phi(R_e))}$. 
For a given (sufficiently high) temperature, 
there are two possible hairy black hole masses that can be in equilibrium with 
thermal radiation. The lower of these has negative specific heat so it is 
unstable to decay into pure thermal radiation. The higher mass black hole, by 
contrast, has always positive specific heat
(and smaller action) so we concentrate on this.

For the asymptotic
conditions on the fields defined in Section 2, both the gravitational
and scalar sector give divergent contributions to the variation of the
boundary term ({\ref{varsurf}) at infinity. However, it was shown in 
\cite{Hertog04} (and in a Euclidean setting in 
\cite{Martinez04,Hertog04c}) that the divergences
cancel out, yielding a finite total surface term at infinity,
\be \label{surfinfty}
B|_\infty=\beta E_h. 
\ee
At the horizon we have
\be 
\delta B|_{R_e}=4\pi\beta e^{-\delta(R_e)} 
R_e\delta f|_{R_e}. 
\ee
Using
\be 
\delta f|_{R_e}=-f'(R_e)\delta R_e  
\ee
this gives, upon integration,
\be\label{surfhor}
B|_{R_e}=-4\pi
\int (\beta e^{-\delta(R_e)}f(R_e)) R_e\delta R_e
=-16\pi^2\int R_e\delta R_e=-8\pi^2R_e^2, 
\ee 
where $\delta R$ is the variation along the horizon. 
Combining (\ref{surfinfty}) and (\ref{surfhor}) and taking into account
the Hamiltonian constraint ${\cal H}=0$, one obtains 
\be \label{acth}
I_h=\beta E_h -8\pi^2 R_h^2 =\beta(E_h-ST)=\beta F,  
\ee
where $S=\pi R_e^2/G$ and $F$ are the entropy and 
free energy, respectively.  

On the other hand, for the Euclidean Schwarzschild AdS black hole we have
\be
\delta =0, \quad f=1-\frac{E_s}{4\pi r}+r^2, \quad R=r.
\ee
The temperature is related to the horizon radius $R_s$ in the following way,
\be
2\pi T = \frac{1 + 3R_s^2}{2R_s}.
\ee
Therefore, the radius of the Schwarzschild-AdS black hole 
in the canonical ensemble is smaller compared to the horizon size of 
the hairy black hole. The Euclidean action is given by
\be \label{acts}
I_s=4\pi(\beta (R_s +R_s^3)-2\pi R_s^2).
\ee

\begin{figure}[htb]
\begin{picture}(0,0)
\put(46,254){$\frac{\Delta I}{2\pi}$}
\put(412,23){$2\pi T$}
\end{picture}
\mbox{\epsfxsize=14cm \epsfysize=9cm \epsffile{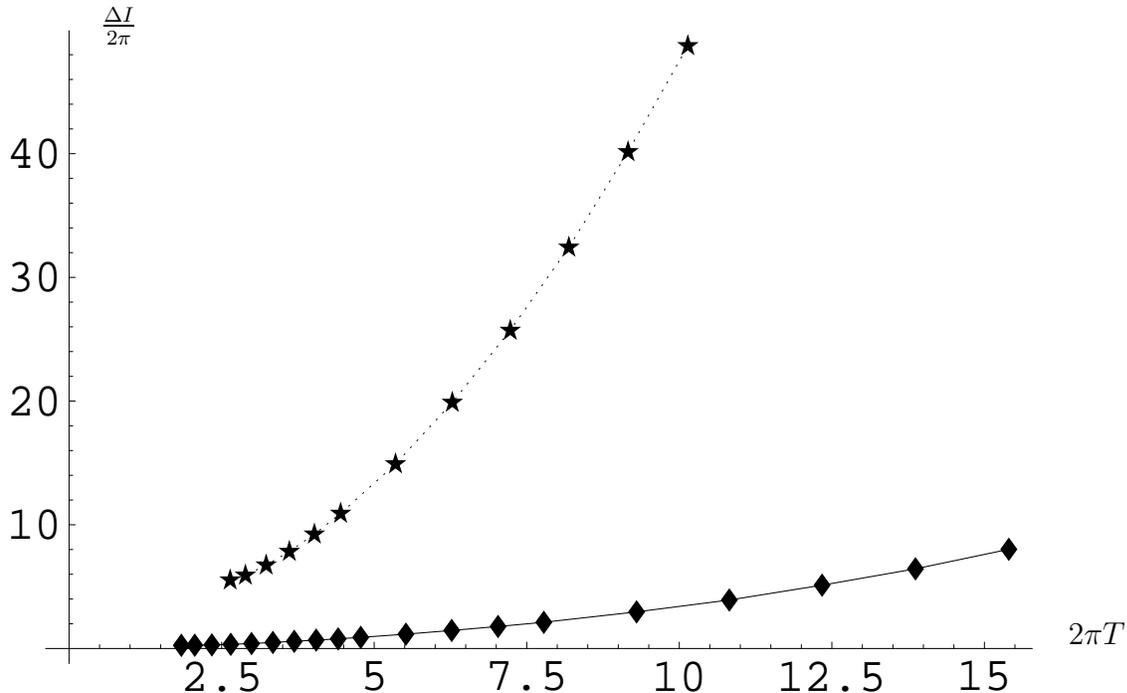}}
\caption{The difference $\Delta I$ between the Euclidean action of 
a hairy black hole, $I_h$, and that of Schwarzschild-AdS, $I_s$, 
as a function of 
temperature $T$. The two curves correspond to hairy black hole solutions 
of $D=4$ gauged supergravity with two different AdS invariant 
boundary conditions, specified by $c=-1$ (bottom) and $c=-1/4$ (top).
One sees that in both cases Schwarzschild-AdS is always 
thermodynamically favored.}
\label{4}
\end{figure}

Using the results of \cite{Hertog04} we have numerically computed the 
difference between the Euclidean actions (\ref{acth}) and (\ref{acts}),
\be
\Delta I = I_h - I_s,
\ee
as a function of temperature\footnote{The curve in the $c=-1/4$ theory starts
at higher temperature because the hairy 
black hole solution does not contribute to the thermodynamic ensemble at 
temperatures below $T= \sqrt{\vert V(R_e)\vert }$.} and in two different
theories. The result is shown
in Figure 4. One sees that the action of the hairy black hole is always larger
than that of Schwarzschild-AdS. This means the latter is thermodynamically 
stable, despite the fact there is no Positive Mass theorem for the set of
asymptotic conditions we considered here. 

We note that our results differ from those obtained in \cite{Martinez04},
where a thermodynamical analysis was performed in which the action of 
asymptotically locally AdS hairy black holes was compared with the action of
vacuum black holes with the same asymptotic structure. The important point is 
that with locally AdS boundary conditions, there is no critical temperature 
below which black holes don't contribute to the thermodynamic ensemble. 
In this context, \cite{Martinez04} finds indeed that for sufficiently low 
temperatures, the vacuum black hole is likely to decay into a hairy 
solution.

\setcounter{equation}{0}
\section{Hairy Black Holes in $D=5$ Supergravity} 

Finally, we briefly consider five dimensional gravity minimally coupled to a scalar field 
with action
\be \label{act5d}
S=\int d^5x\sqrt{-g}\left[\frac{1}{2}R-\frac{1}{2}(\nabla\phi)^2 
+\left(2e^{2\phi /\sqrt{3}} +4e^{-\phi/\sqrt{3}}\right)\right].
\ee
This is a consistent truncation of ${\cal N}=8$ gauged supergravity in five
dimensions, which is the low energy limit of string theory with 
$AdS_5 \times S^5$ boundary conditions. At the global negative maximum of the potential
at $\phi=0$, the scalar has mass $m^2=-4$ and thus saturates the Breitenlohner-Freedman 
bound in five dimensions.

In \cite{Hertog04} a class of static, spherically symmetric black holes was found 
with a nontrivial scalar field profile outside the horizon. Asymptotically, the scalar
field behaves as
\be 
\label{5-scalar}
\phi (r)=\frac{\alpha_0}{r^2}\ln r+\frac{\alpha_0}{r^2} \left(c-\frac{1}{2} \ln \alpha_0 \right)
\ee
where $c$ is a constant that labels the choice of boundary conditions and
$\alpha_0$ depends on the horizon size. The backreaction of
the scalar field causes the $g_{rr}$ component of the metric to fall off slower than usual,
but the asymptotic conditions are left invariant under the full AdS symmetry group.
The conserved charges generating the symmetries are finite, but acquire a 
contribution from the scalar field \cite{Hertog04,Henneaux04}. 
In particular, the mass of the hairy black holes is given by
\be \label{mass5dhair}
E_{h} =2\pi^2  \left(\frac{3}{2} M_0+\frac{1}{4}\alpha^2_0 (\ln \alpha_0 )^2 +\alpha^2_0 
\left( \frac{1}{4} -c \right) \ln \alpha_0 +\alpha^2_0 \left( c^2 -\frac{1}{2}c +\frac{1}{8} \right) \right)
\ee
where $M_0$ is now the ${\cal O}(r^{-6})$ correction to $g_{rr}$.
It is interesting to verify  the stability of these hairy black holes, 
especially since
$D=5$, ${\cal N} =8$ supergravity plays a prominent role in the AdS/CFT
correspondence. Substituting the following expansions 
\beq
\phi(r,t)&=&\phi_0(r)+\frac{\chi(r)}{r^{3/2}}
e^{i\omega t}\eta,  \nonumber \\
M(r,t)&=&M_0(r)+M_1(r)e^{i\omega t}\eta, \nonumber \\
\delta(r,t)&=&\delta_0(r)+\delta_1(r)e^{i\omega t}\eta
\eeq
in the field equations one obtains, to first order in $\eta$, 
an equation for the fluctuations of the form (\ref{perteq}), with 
\be \label{perteq5}
P(r)=e^{-2\delta_0}f_0
\left[f_0\left(\frac{3}{4r^2}-\frac{5\phi_0'^2}{6}
-\frac{2r^2}{9}\phi_0'^4\right)
+f'_0\left(\frac{3}{2r}-\frac{2r}{3}\phi_0'^2\right)
+\frac{4r}{3}\phi_0' V_{,\phi}(\phi_0)
+V_{,\phi \phi}(\phi_0)\right].
\ee 
The potential is unbounded from below, since
the scalar mass $m^2$ is less than the conformal mass $m^2_c$. 
Near spacelike infinity $P$ goes as
\be
P \rightarrow -\frac{1}{4}e^{-2\delta_0}r^2,
\ee
so one expects there should exist an unstable mode.
The general asymptotic solution of (\ref{perteq5}) reads
\be 
\chi=\frac{\alpha_1}{\sqrt{r}} \ln r +\frac{\beta_1}{\sqrt{r}} 
\ee
This belongs to the same class of asymptotically AdS spacetimes as the
black hole background itself, provided that
\be 
\frac{\beta_1}{\alpha_1}=\left( -\frac{1}{2} \ln \alpha_0+c-\frac{1}{2}
\right). 
\ee
We find there is always precisely one negative frequency $\omega^2_{n}$ for 
which
this condition is satisfied. For instance, for a black hole of radius 
$R_e =1$ in the $c=0$ theory, the unstable mode has 
frequency $\omega^2_{n} =-2.4$.

\setcounter{equation}{0}
\section{Discussion}

We have studied the thermodynamics and stability of a class of 
spherical hairy black holes, which are solutions
of $D=4$ gauged supergravity with modified, but AdS invariant boundary
conditions. 
The thermodynamical analysis shows that in the canonical ensemble the usual
Schwarzschild-AdS black hole is stable against `decay' into a hairy black hole.
The stability analysis reveals there always exists precisely one
unstable radial fluctuation around the hairy black hole background.
We find similar results in $D=5$ supergravity.

We should note that 
the instability of the supergravity hairy black hole solutions is not only
a consequence of the modification of the standard AdS boundary
conditions, which renders the AdS state itself (nonlinearly) unstable.
Rather, it is because the scalar field mass $m^2$ is  
less than or equal to the conformal mass $m^2_c$ in both cases. 
Stable hairy AdS black hole exist in theories where $m^2_c <m^2 <m^2_{BF}+1$ 
\cite{Torii:2001pg}\footnote{Although the conventional mass of the 
solutions presented in \cite{Torii:2001pg} diverges, the example with 
$\alpha=1.55$ in the symmetric double well potential can be reinterpreted as
an asymptotically AdS black hole obeying the modified boundary conditions
given in \cite{Hertog04}.}. 
It would be interesting to find a supergravity example of this.

We can argue that Schwarzschild-AdS is unlikely to be the endstate 
of the evolution of a perturbed hairy black hole in the following 
way. Assume for a moment the perturbed hairy black hole does tend to 
Schwarzschild-AdS and that the decay of the scalar field near the horizon
is governed by the lowest quasinormal mode. 
Then one could approximate the evolution 
of the scalar field $\phi$ near the horizon as
\be
\label{phi-as}
\phi \simeq \phi_e e^{-2\pi \gamma T (v-v_0)},   
\ee
where $\gamma$ is a constant and $v=t+r^*$. 
Here we used the fact that the imaginary part of the frequency of 
quasinormal modes is proportional to the temperature $T$ of Schwarzschild-AdS
\cite{Horowitz00}. Now consider a null geodesic with 
tangent vector $V^\mu$ on the event horizon at radius $R$. 
Since the affine parameter 
$\lambda$ is related to 
$v$ as 
\be 
\lambda\sim \lambda_0e^{2\pi T(v-v_0)}, 
\ee
eq.~(\ref{phi-as}) becomes 
\be 
\phi\sim \phi_e\left(\frac{\lambda}{\lambda_0}\right)^{-\gamma}.  
\ee
On the other hand, the Raychaudhuri equation reads
\be 
\frac{d\theta}{d\lambda}=-\frac{1}{2}\theta^2-T_{\mu\nu}V^\mu V^\nu.   
\ee
Using 
\be
T_{\mu\nu}V^\mu V^\nu=\dot{\phi}^2, \qquad \theta=\frac{2\dot{R}}{R}, 
\ee
this becomes
\be \label{raych}
\ddot{R}=-\frac{\dot{\phi}^2}{2}R, 
\ee
where $\dot \phi =\partial_{\lambda}\phi$. 
Since the perturbed hairy black hole must tend to Schwarzschild-AdS 
of the same total mass, we have (for large initial horizon size $R_e$)
\be     
E_h=4\pi R_e^3(1+{\cal O}(1/R_e)) =4\pi(R_f^3+R_f),
\ee
where $R_f$ is the radius of the final Schwarzschild-AdS configuration. 
This means that the final radius $R_f$ cannot be a multiple of $R_e$, but
instead $R_f =R_e+c_0$, where $c_0$ is a positive constant.  
Hence (\ref{raych}) can be approximated as follows,
\be 
\ddot{R}\simeq -\frac{\dot{\phi}^2}{2}R_e=
-\frac{R_e}{2}\left(\frac{\phi_e \gamma}{\lambda_0}\right)^2 
\left(\frac{\lambda}{\lambda_0}\right)^{-2\gamma-2}.    
\ee
Integrating yields,
\be 
R_f=R(\lambda=\infty)=R_e\left(1+\frac{\phi_e^2 \gamma}{2(2\gamma+1)}\right). 
\ee
But this contradicts the fact that 
$R_f=R_e+c_0$, since for instance in the $c=-1$ theory one has
$\phi_e \rightarrow 0.25$ for large $R_e$. 

This indicates that Schwarzschild-AdS is probably not
the endstate of evolution under the instability. Instead, the perturbed 
black hole is likely to continue to evolve forever, perhaps in a manner
envisioned in \cite{Alcubierre04} (in a different context)
where the horizon continues to expand.
This raises the question whether or not the black hole singularity reaches 
the boundary of AdS in finite time. If so, small fluctuations would 
induce a transition between a black hole and a big crunch.
The surprising evolution under the instability can also be understood from 
the dual CFT perspective, where the hair turns on the triple trace operator
\cite{Hertog04c}. This indicates 
that, in contrast to Schwarschild-AdS black holes, the hairy black holes are 
states concentrated on the unstable side 
of the field theory potential. The evolution of some states of this kind has 
been shown before to produce a big crunch singularity \cite{Hertog04c}.
One can then hope to use AdS/CFT to shed light
on the quantum nature of the spacelike singularities in these dynamical
solutions.

\bigskip

\centerline{{\bf Acknowledgments}}

We thank T. Torii for helpful discussions.
This work was supported in part by NSF grant PHY-0244764.

\end{document}